\documentclass[preprint,english,showpacs,11pt,floatfix]{revtex4}

\usepackage{babel,amsmath,amssymb,dcolumn}
\usepackage[dvips]{graphics}

\begin{document}

\title{Quasinormal modes in time-dependent black hole background}

\author{Cheng-Gang Shao, Bin Wang}
\email{binwang@fudan.ac.cn} \affiliation{Department of Physics,
Fudan University, Shanghai 200433, People's Republic of China }

\author{Elcio Abdalla}
\email{eabdalla@fma.if.usp.br} \affiliation{Instituto de Fisica,
Universidade de Sao Paulo, C.P.66.318, CEP 05315-970, Sao Paulo,
Brazil}

\author{Ru-Keng Su}
\email{rksu@fudan.ac.cn} \affiliation{China Center of Advanced
Science and Technology (World Laboratory), P.B.Box 8730, Beijing
100080, People's Republic of China
\\Department of Physics, Fudan University, Shanghai 200433,
People's Republic of China }

\begin{abstract}
We have studied the evolution of the massless scalar field
propagating in time-dependent charged Vaidya black hole
background. A generalized tortoise coordinate transformation were
used to study the evolution of the massless scalar field. It is
shown that, for the slowest damped quasinormal modes, the
approximate formulae in stationary Reissner-Nordstr\"{o}m black
hole turn out to be a reasonable prescription, showing that
results from quasinormal mode analysis are rather robust.
\end{abstract}


\maketitle

\section{Introduction}

It is well known that the surrounding geometry of a black hole
will experience damped oscillations under perturbations. These
oscillations are called "quasinormal modes" (QNM), which is
believed to be a characteristic "sound" of black holes, and would
lead to the direct identification of the black hole existence
through gravitational wave observation, to be realized in the near
future. In the past few decades, a great deal of effort has been
devoted to the study of QNMs of black holes in asymptotically flat
spacetimes (for comprehensive reviews see [1,2] and references
therein). Considering the case when the black hole is immersed in
an expanding universe, the QNMs of black holes in de Sitter space
have also been investigated [3-5]. Motivated by the recent
discovery of the Anti-de Sitter/Conformal Field Theory (AdS/CFT)
correspondence, many authors have performed the study of QNMs in
AdS spaces [6-18]. The study of the QNMs plays an important role
in astrophysics, black hole physics and also string theory.

All these previous works on QNMs have so far been restricted to
time-independent black hole backgrounds. It should be realized
that, for a realistic model, the black hole parameters change with
time. A black hole gaining or losing mass via absorption (merging)
or evaporation is a good example. The more intriguing
investigation of the black hole QNM calls for a systematic
analysis of time-dependent spacetimes. Recently the late time
tails under the influence of a time-dependent scattering potential
has been explored in [19], where the tail structure was found to
be modified due to the temporal dependence of the potential. The
motivation of our work is to explore the modification to the QNM
in time-dependent spacetimes. Instead of plotting an effective
time-dependent scattering potential by hand as done in [19], we
will introduce the time-dependent potential in a natural way by
considering dynamical black holes, with black hole parameters
changing with time due to absorption and evaporation processes. We
will study the temporal evolution of massless scalar field
perturbation. Instead of employing Kruskal-like coordinate in our
first attempt before [20], here we start our discussion directly
from the Vaidya metric.

The outline of this paper is as follows. In Sec. II we first go
over the conventional treatment for the study of the wave
propagation in stationary Reissner-Nordstr\"{o}m (RN) black hole.
In Sec. III, we will present the numerical way for the study of
the QNMs of charged Vaidya black hole. We will present our
numerical results in Sec. IV. Sec. V contains a brief summary and
a discussion.

\section{Quasinornal modes of a stationary Reissner-Nordstr\"{o}m black hole}

Perturbation outside the stationary RN black hole has been
discussed by many authors [21]. Here we first present a brief
review on the conventional treatment of the evolution of massless
scalar field propagating in RN black hole background.

The metric of a RN black hole in the ingoing Eddington coordinates
is given by
\begin{equation}
\label{eq1} ds^2 = - (1 - 2M / r + Q^2 / r^2)dv^2 + 2dvdr +
r^2(d\theta ^2 + \sin ^2\theta d\varphi ^2) \equiv g_{\mu \nu }
dx^\mu dx^\nu ,
\end{equation}
where $v$ is a coordinate usually called the `advanced time',
which parameterizes a lapse of time. The propagation of scalar
waves in curved spacetimes is governed by the Klein-Gordon
equation
\begin{equation}
\label{eq2}
\partial _\mu (\sqrt { - g} g^{\mu \nu }\partial _v )\Phi = 0 .
\end{equation}
Since the background is spherically symmetric, each multipole of
the perturbing field evolves separately. Hence one can define
\begin{equation}
\label{eq3} \Phi = \sum\limits_{l,m} {\Psi (r,v)Y_{lm} (\theta
,\varphi ) / r} .
\end{equation}

Using the tortoise coordinate $r_\ast $ defined by
\begin{equation}
\label{eq4} r_\ast = r + \frac{r_ + ^2 }{r_ + - r_ - }\ln (r - r_
+ ) - \frac{r_ - ^2 }{r_ + - r_ - }\ln (r - r_ - ) ,
\end{equation}
where, $r_ + = M + \sqrt {M^2 - Q^2} $ and $r_ - = M - \sqrt {M^2
- Q^2} $ are the radius of outer and inner horizons of the black
hole, the wave equation for each multipole moment becomes
\begin{equation}
\label{eq5} \Psi _{,r_\ast r_\ast } + 2\Psi _{,r_\ast v} - V\Psi =
0 ,
\end{equation}
where
\begin{equation}
\label{eq6} V(r_\ast ) = (1 - \frac{2M}{r} +
\frac{Q^2}{r^2})(\frac{l(l + 1)}{r^2} + \frac{2M}{r^3} -
\frac{2Q^2}{r^4}) .
\end{equation}

Using the null coordinate
\begin{equation}
\label{eq7} u = v - 2r_\ast
\end{equation}
equation (\ref{eq5}) can be written as
\begin{equation}
\label{eq8} \Psi _{,uv} + \frac{1}{4}V\Psi = 0 .
\end{equation}
The two-dimensional wave equation (\ref{eq8}) can be integrated
numerically, using for example the finite difference method
suggested in [22]. Using Taylor's theorem, it is discretized as
\begin{equation}
\label{eq9} \Psi _N = \Psi _E + \Psi _W - \Psi _S - \delta u\delta
vV(\frac{v_N + v_W - u_N - u_E }{4})\frac{\Psi _W + \Psi _E }{8} +
O(\varepsilon ^4) ,
\end{equation}
where the points $N$, $S$, $E$ and $W$ from a null rectangle with
relative positions as: $N:(u + \delta u,v + \delta v)$,$W:(u +
\delta u,v)$,$E:(u,v + \delta v)$ and $S:(u,v)$. The parameter
$\varepsilon $ is an overall grid scalar factor, so that $\delta
u\sim \delta v\sim \varepsilon $. For an RN black hole in de
Sitter space this has been performed in [4].

In addition to solving (\ref{eq8}) directly by numerical method,
other ways have been suggested in studying the QNMs of RN black
hole. For the slowest damped QNMs, the WKB formulas suggest the
approximate behavior in RN black hole [23]
\begin{equation}
\label{eq10}
\begin{array}{l}
 \omega _R \approx (l + \frac{1}{2})\left[ {\frac{M}{r_0^3 } -
\frac{Q^2}{r_0^4 }} \right]^{1 / 2} , \\
 \omega _I \approx - \frac{1}{2}\left[ {\frac{M}{r_0^3 } - \frac{Q^2}{r_0^4
}} \right]^{1 / 2}\left[ {2 - 3\frac{M}{r_0 }} \right]^{1 / 2} ,\\
 \end{array}
\end{equation}
in the limit $l > > 1$. Here $r_0 = \frac{3}{2}M +
\frac{1}{2}(9M^2 - 8Q^2)^{1 / 2}$ is the position where the
potential $V$ attains its maximum value. Defining $q = Q / M $,
the above equation becomes
\begin{equation}
\label{eq11}
\begin{array}{l}
 M\omega _R \approx c_R (l,q) \equiv (l + \frac{1}{2})\left[ {\frac{3}{2}(1
+ \sqrt {1 - \frac{8}{9}q^2} ) - q^2} \right]^{1 / 2}\left[ {\frac{9}{2} -
2q^2 + \frac{9}{2}\sqrt {1 - \frac{8}{9}q^2} } \right]^{ - 1} ,\\
 M\omega _I \approx c_I (l,q) \equiv - \frac{3}{2}\left[ {\frac{3}{2}(1 +
\sqrt {1 - \frac{8}{9}q^2} ) - q^2} \right]^{1 / 2}\left[ {\frac{3}{2}(1 +
\sqrt {1 - \frac{8}{9}q^2} )} \right]^{ - 3}\sqrt {1 - \frac{8}{9}q^2} .\\
\end{array}
\end{equation}
The WKB computation has also been performed in the RN de Sitter
Spacetime [4].

These conventional treatments are powerful in studying the wave
propagation in stationary black hole background, however it is
difficult to extend  it to the time-dependent case. For the
charged Vaidya black hole, the mass and the charge are functions
of time, $M = M(v)$,$Q = Q(v)$, and the standard tortoise
coordinate can not be used to simplify the wave equation.

\section{Field evolution in charged Vaidya background}

The metric of the charged Vaidya spacetime reads
\begin{equation}
\label{eq12} ds^2 = - (1 - \frac{2M(v)}{r} +
\frac{Q^2(v)}{r^2})dv^2 + 2c dvdr + r^2(d\theta ^2 + \sin ^2\theta
d\varphi ^2) ,
\end{equation}
where $M = M(v)$ and $Q = Q(v)$ are arbitrary functions of the
time. For $c=1$ the field is ingoing and $M$ is monotone
increasing in $v$ (advanced time), and for $c=-1$, the field is
outgoing and $M$ is monotone decreasing in $v$ (retarded time)
[24]. The QNMs of a black hole correspond to the solutions of the
Klein-Gordon equation that satisfy the causal condition that no
information could leak out through the event horizon of the black
hole and at the same time correspond to purely outgoing waves at
spatial infinity. For the charged Vaidya black hole, horizons
$r_\pm $ can be inferred from the null hypersurface condition
$g^{\mu \nu }f_{,\mu } f_{,\nu } = 0$ and $f(r_\pm ,v) = 0$.
$r_\pm (v)$ satisfies the equation
\begin{equation}
\label{eq13} r_\pm ^2 - 2r_\pm M + Q^2 - 2c r_\pm ^2 \dot {r}_\pm
= 0 ,
\end{equation}
where $\dot {r}_\pm \equiv dr_\pm / dv$.

Since the charged Vaidya metric is spherically symmetric, a
generalized tortoise coordinate transformation can be introduced
as
\begin{equation}
\label{eq14} r_\ast = r + \frac{1}{2k_ + }\ln (r - r_ + ) -
\frac{1}{2k_ - }\ln (r - r_ - ),v_\ast = v - v_0 ,
\end{equation}
which is similar to equation (\ref{eq4}), and which we take as a
simplifying Ansatz for our numerical equations upon an
appropriated choice of the constants $k_\pm $, as well as the
functions $r_\pm (v)$, which will be interpreted as the event and
Cauchy horizons \footnote{Actually, the event horizon is a global
notion in the Vadyia Metric (see [26]). We nevertheless borrow
here this terminology.}, $r_ + = M + \sqrt {M^2 - Q^2 + 2c r_ + ^2
\dot {r}_ + } $ and $r_ - = M - \sqrt {M^2 - Q^2 + 2c r_ - ^2 \dot
{r}_ - } $, respectively. We take $k_ + $ and $k_ - $ as
adjustable parameters to be defined. The parameter $v_0 $ is an
arbitrary constant assumed to be zero now. From formulas
(\ref{eq14}) and (\ref{eq12}), the Klein-Gordon equation for each
multipole moment becomes
\begin{equation}
\label{eq15} (1 + \varepsilon _2 )\Psi _{,r_\ast r_\ast } + 2c
\Psi _{,r_\ast v_\ast } + \varepsilon _1 \Psi _{,r_\ast } - V\Psi
= 0 ,
\end{equation}
where
\[
\varepsilon _2 = \frac{1}{r}\left( {\frac{1 - 2c \dot {r}_ + }{2k_
+ } - \frac{1 - 2c \dot {r}_ - }{2k_ - } - 2M} \right) +
\frac{1}{r^2}\left( {\frac{(1 - 2c \dot {r}_ + )r_ + - 2M}{2k_ + }
- \frac{(1 - 2c \dot {r}_ - )r_ - - 2M}{2k_ - } + Q^2} \right) ,
\]

\[
\varepsilon _1 = A\left( {\frac{1}{r^2}(\frac{1 - 2c \dot {r}_ +
}{2k_ + } - \frac{1 - 2c \dot {r}_ - }{2k_ - } - 2M) -
\frac{2\varepsilon _2 }{r}} \right) ,
\]

\begin{equation}
\label{eq16}
\begin{array}{l}
 V = A\frac{l(l + 1)}{r^2} + A^2\left( {\frac{1}{r^3}(\frac{1 - 2c\dot {r}_ +
}{2k_ + } - \frac{1 - 2c\dot {r}_ - }{2k_ - } - 2M) -
\frac{2\varepsilon _2 }{r^2}} \right) - 2c A^2\frac{1}{r}\left(
{\frac{\dot {r}_ + }{2k_ + (r - r_ +
)^2} - \frac{\dot {r}_ - }{2k_ - (r - r_ - )^2}} \right) \\
 {\begin{array}{*{20}c}
 \hfill & \hfill \\
\end{array} } + 2A^3\frac{1}{r}\left( {\frac{1}{2k_ + (r - r_ + )^2} -
\frac{1}{2k_ - (r - r_ - )^2}} \right)\left( {\frac{c \dot {r}_ +
}{2k_ + (r - r_ + )} - \frac{c \dot {r}_ - }{2k_ - (r - r_ - )} +
\frac{1 + \varepsilon _2
}{2}} \right) ,\\
 \end{array}
\end{equation}
with
\[
A = \left( {1 + \frac{1}{2k_ + (r - r_ + )} - \frac{1}{2k_ - (r -
r_ - )}} \right)^{ - 1} .
\]

For the convenience of numerical calculation, the adjustable
parameters $k_ + $ and $k_ - $ can be selected in such a way that
the Klein-Gordon equation for each multipole moment becomes the
standard wave equation near the horizons $r_ + (v_0 )$ and $r_ -
(v_0 )$. We define $k_ + $ and $k_ - $ by the algebraic equations
\begin{equation}
\label{eq17}
\begin{array}{l}
 \frac{r_ - - M - 2c r_ - \dot {r}_ - }{r_ - ^2 }\frac{1}{k_ - } - c \frac{\dot
{r}_ + - \dot {r}_ - }{r_ + - r_ - }\frac{1}{k_ + } = - 1 + 2c \dot {r}_ -  ,\\
 \frac{r_ + - M - 2c r_ + \dot {r}_ + }{r_ + ^2 }\frac{1}{k_ + } - c \frac{\dot
{r}_ + - \dot {r}_ - }{r_ + - r_ - }\frac{1}{k_ - } = 1 - 2c \dot {r}_ +  ,\\
 \end{array}
\end{equation}
at $v = v_0 $. We impose for simplicity, that
\begin{equation}
\label{eq18} \mathop {\lim }\limits_{\begin{array}{l}
 r \to r_ \pm (v_0 ) \\
 v \to v_0 \\
 \end{array}} \varepsilon _2 = \mathop {\lim }\limits_{\begin{array}{l}
 r \to r_ \pm (v_0 ) \\
 v \to v_0 \\
 \end{array}} \varepsilon _1 = \mathop {\lim }\limits_{\begin{array}{l}
 r \to r_ \pm (v_0 ) \\
 v \to v_0 \\
 \end{array}} V = 0 .
\end{equation}
For the static black hole with $\dot {r}_ + = \dot {r}_ - = 0$ (or
$M$ and Q are constants), we have the usual result
$\frac{1}{2k_\pm } = \frac{r_\pm ^2 }{r_ + - r_ - }$, when
(\ref{eq14}) boils down to (\ref{eq4}) and (\ref{eq15}) to
(\ref{eq5}).

Analogous to the coordinate used in (\ref{eq4}), we make the
variable transformation
\begin{equation}
\label{eq19} u = u(r_\ast ,v_\ast ),v = v_\ast
\end{equation}
where, the curve $u(r_\ast ,v_\ast ) = $constant is determined by
the equation
\begin{equation}
\label{eq20} \frac{dr_\ast }{dv_\ast } = \frac{1 + \varepsilon _2
}{2c} .
\end{equation}
For $\varepsilon _2 \to 0$ and $c=1$, we have $u \to v_\ast -
2r_\ast $, which is similar to (\ref{eq7}). Using the
transformation (\ref{eq19}), equation (\ref{eq15}) can be
simplified to
\begin{equation}
\label{eq21} \Psi _{,uv} - (2c \frac{\partial u}{\partial r_\ast
})^{ - 1}V\Psi = 0 .
\end{equation}
The function $u(r_\ast ,v_\ast )$ can be integrated numerically
according to equation (\ref{eq20}). The integration of $\Psi $ can
proceed similarly to equation (\ref{eq9}).

However, the numerical calculation discussed above has some
difficulty for small $Q$, especially for the Vaidya metric with $Q
\to 0$. In this case the inner horizon $r_ - $ disappears and the
transformation (\ref{eq14}) is no longer valid. In this case
(\ref{eq14}) can be replaced by [25]
\begin{equation}
\label{eq22} r_\ast = r + \frac{1}{2k}\ln (r - r_ +
),{\begin{array}{*{20}c}
 \hfill & \hfill \\
\end{array} }v_\ast = v - v_0
\end{equation}
where $k$ is an adjustable parameter. Using (\ref{eq22}), the
Klein-Gordon equation for each multipole moment has the same form
as equation (\ref{eq15}), only replacing the expressions of
$\varepsilon _2 $, $\varepsilon _1 $ and V by
\[
\varepsilon _2 = [2k(Q^2 - 2Mr) + r + r_ + - 2M - 2c(r + r_ +
)\dot {r}_ + ] / (2kr^2) ,
\]

\[
\varepsilon _1 = \frac{2k(r - r_ + )}{2k(r - r_ + ) + 1}\left(
{\frac{1 - 2c \dot {r}_ + - 4kM}{2kr^2} - \frac{2\varepsilon _2
}{r}} \right) ,
\]

\begin{equation}
\label{eq23} V = \frac{2k(r - r_ + )}{2k(r - r_ + ) + 1}\left(
{\frac{\varepsilon _1 }{r} + \frac{2k(1 - 2c \dot {r}_ + ) +
2k\varepsilon _2 }{(2k(r - r_ + ) + 1)^2r} + \frac{l(l + 1)}{r^2}}
\right) .
\end{equation}
The parameter $k$ can be selected to simplify the Klein-Gordon
equation. We choose it such that
\begin{equation}
\label{eq24} \mathop {\lim }\limits_{\begin{array}{l}
 r \to r_ + (v_0 ) \\
 v \to v_0 \\
 \end{array}} \varepsilon _2 = \mathop {\lim }\limits_{\begin{array}{l}
 r \to r_ + (v_0 ) \\
 v \to v_0 \\
 \end{array}} \varepsilon _1 = \mathop {\lim }\limits_{\begin{array}{l}
 r \to r_ + (v_0 ) \\
 v \to v_0 \\
 \end{array}} V = 0 ,
\end{equation}
namely
\begin{equation}
\label{eq25} k = \frac{r_ + (v_0 )M(v_0 ) - Q(v_0 )^2}{[2r_ + (v_0
)M(v_0 ) - Q(v_0 )^2]r_ + (v_0 )} .
\end{equation}
In the numerical calculation for the field evolution $\Psi $, the
coordinate transformation (\ref{eq22}) can be appropriated for $Q
\to 0$.

In the following numerical calculation for the field $\Psi $, we
always use a $\delta $ pulse as an initial perturbation, located
at $v_0 = 0,r = 3$. We first use our numerical method to the case
of stationary RN black hole. The numerical result is in agreement
with equation (\ref{eq10}) or (\ref{eq11}), even for $l$=2. Table
1 shows the slowest damped quasinormal frequencies for the scalar
perturbation in the RN background with the multipole index $l =
2$. The numerical result is also in agreement with Ref. [4] for
$q$=0.

\begin{table}[htbp]
\caption{The numerical result for $M\omega _R $ (or $M\omega _I )$
and the lowest WKB approximate formulas for $c_R (2,q)$ (or $c_I
(2,q))$ in the slowest damped quasinormal frequencies in the RN
background. The multipole index $l = 2$.}
\begin{tabular}
{|p{45pt}|p{45pt}|p{45pt}|p{45pt}|p{45pt}|p{45pt}|p{45pt}|} \hline
q& $M\omega _R $& $c_R (2,q)$& $M\omega _I $& $c_I (2,q)$&
$M\omega _R $ [4]& $M\omega _I $ [4] \\
\hline 0&0.483& 0.481& -0.0965& -0.0962& 0.484&-0.0965 \\
\hline 0.7& 0.532& 0.530& -0.0985& -0.0981& & \\
\hline 0.999& 0.626& 0.624& -0.0889& -0.0886& & \\
\hline
\end{tabular}
\label{tab1}
\end{table}

\section{Numerical result of QNMs in the charged Vaidya black hole}

It is widely believed today that asymptotically flat black holes
eventually evaporate due to the emission of Hawking radiation. For
a RN black hole, we assume that at a certain retarded moment $v =
v_0 $ a charged nullfluid of negative-energy-density starts
falling into the black hole (corresponding to the beginning of the
evaporation process). We first adopt the `linear model' [26],
where $M$ and $Q$ depend linearly on $v$, and we also assume that
the charge to mass ratio remains fixed.
\begin{equation}
\label{eq26}
\begin{array}{l}
 M(v) = \left\{ {{\begin{array}{l l}
 {m_0 } & {v \le v_0 } \\
 {m_0 (1 - \lambda v)} & {v_0 \le v \le v_1 } \\
 {m_1 } & {v \ge v_1 } \\
\end{array} }} \right. \\
 Q(v) = qM(v) ,\\
 \end{array}
\end{equation}
where $m_0 $, $m_1 $, $\lambda $ and $q$ are constant parameters.
The above model is initially a RN black hole for $v < v_0 $. The
evaporation process is displayed in the region $v_0 < v < v_1 $.
At $v = v_1 $, the evaporation process stops, and the final
geometry $v > v_1 $ is again the RN black hole. Changing the sign
before $\lambda$ in (\ref{eq26}), this model can mimic the black
hole absorption process. In the numerical study, we only consider
how does the absorption or evaporation process affect the QNMs of
the black hole in the region $v_0 < v < v_1 $.

For the linear model (\ref{eq26}), we can get the solution $r_\pm
$ of equation (\ref{eq13}) as
\begin{equation}
\label{eq27}
\begin{array}{l l}
 r_ + = M + M\sqrt {1 - q^2} & v > v_1 ,\\
 r_ + = \frac{M + M\sqrt {1 - q^2(1 - 2\dot {r}_ + )} }{1 - 2\dot {r}_ +
}, & v \le v_1 ,\\
 r_ - = \frac{M - M\sqrt {1 - q^2(1 - 2\dot {r}_ + )} }{1 - 2\dot {r}_ +
}, & v \ge v_0 ,\\
 r_ - = M - M\sqrt {1 - q^2} & v < v_0 ,\\
 \end{array}
\end{equation}
for absorption process with $c=1$, and
\begin{equation}
\label{eq28}
\begin{array}{l l}
  r_ + = \frac{M + M\sqrt {1 - q^2(1 + 2\dot {r}_ + )} }{1 + 2\dot {r}_ +
}, & v \ge v_0 ,\\
r_ + = M + M\sqrt {1 - q^2} & v < v_0 ,\\
r_ - = M - M\sqrt {1 - q^2} & v > v_1 , \\
 r_ - = \frac{M - M\sqrt {1 - q^2(1 + 2\dot {r}_ + )} }{1 + 2\dot {r}_ +
}, & v \le v_1 ,\\
 \end{array}
\end{equation}
for evaporation process with  $c=-1$. In general we suppose that
$\left| \dot {M} \right| = \lambda m_0 < < 1$, which leads to
$r_\pm \approx M\pm M\sqrt {1 - q^2} $. Thus the $r_ + $ is very
close to the outer apparent horizon $M + M\sqrt {1 - q^2} $ [26],
while the location of the $r_ - $ is very close to the inner
apparent horizon $M - M\sqrt {1 - q^2} $. If $q \to 1$, then $r_ +
\approx M$, $\dot {r}_ + \approx \dot {M}$. Meanwhile equations
(\ref{eq27}) and (\ref{eq28}) require $1 - q^2(1 - 2c \dot {r}_ +
) \ge 0$, which is obviously satisfied for both the absorption and
evaporation cases. As an example, we take $m_0 = 0.5$, $\lambda =
0.002$, $v_0 = 0$, $v_1 = 150$ (correspondingly, $m_1 = 0.35$ and
$\dot {M} = - 0.001)$. In Fig.1 we show behaviors of $r_ + $ and
$r_ - $ for three different $q$ in the evaporation process.

\begin{figure}
\resizebox{0.5\linewidth}{!}{\includegraphics*{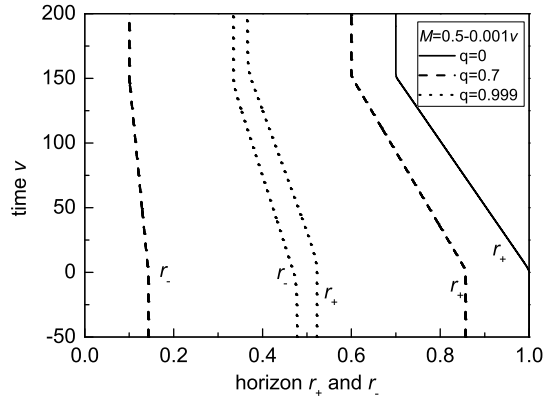}}
\caption{The behavior of $r_ + $ and $r_ - $ for the black hole
with $q$=0, $q$=0.7 and $q$=0.999, respectively. The linear model
is used with $m_0 = 0.5$, $\lambda = 0.002$, $v_0 = 0$, $v_1 =
150$ (correspondingly, $m_1 = 0.35$ and $\dot {M} = - 0.001)$.}
\label{fig1}
\end{figure}

The model (\ref{eq26}) can represent the absorption process of the
black hole when $\lambda < 0$ (or $m_0 < m_1 )$. As an example,
the temporal evolution of the field $\Psi $ in the Vaidya metric
($q$=0) at $r = 5$ is displayed in Fig.2. The relevant parameters
were chosen as $m_0 = 0.5$, $\lambda = \pm 0.002$, $v_0 = 0$,$v_1
= 150$, $l = 2$. The initial $\delta $ pulse perturbation is
located at $v_0 = 0,r = 3$. For comparison, we also exhibit the
curve obtained for Schwarzschild black hole (replacing the
time-dependent mass by a constant mass $M = 0.5)$. The
modification to the QNMs due to the time-dependent background is
clear. When $M$ increases linearly with $v$, the decay becomes
slower compared to the stationary case, which corresponds to
saying that $\left| {\omega _I } \right|$ decreases with respect
to $v$. The oscillation period is no longer a constant as in the
stationary Schwarzschild black hole. It becomes longer with the
increase of time. In other words, the real part of the quasinormal
frequency $\omega _R $ decreases with the increase of time. When
$M$ decreases linearly with respect to $v$, compared to the
stationary Schwarzschild black hole, we have observed that the
decay becomes faster and the oscillation period becomes shorter,
thus both $\left| {\omega _I } \right|$ and $\omega _R $ increase
with time.

\begin{figure}
\resizebox{0.5\linewidth}{!}{\includegraphics*{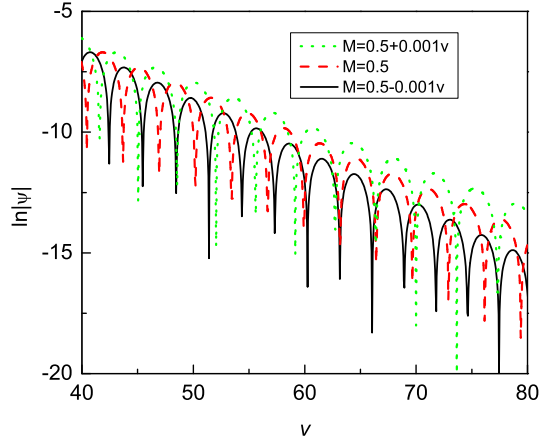}}
\caption{Temporal evolution of the field in the background of
Vaidya metric ($q$=0) for $l = 2$, evaluated at $r = 5$. The mass
of the black hole is $M(v) = 0.5\pm 0.001v$. The field evolution
for $M(v) = 0.5 + 0.001v$ and $M(v) = 0.5 - 0.001v$ are shown as
the top curve and the bottom curve respectively. For comparison,
the oscillations for $M = 0.5$ is given in the middle line.}
\label{fig2}
\end{figure}

Next, we examine the behavior of the QNMs with the increase of the
charge. The Figs. 3-6 display the frequency $\omega _R $ and
$\omega _I $ as a function of $v$ for $l = 2$, evaluated at $r =
5$. The mass of the black hole is $M(v) = 0.5\pm 0.001v$. Both
$\omega _R $ and $\omega _I $ can be determined. Different from
the stationary case, all curves of $\omega _R $ and $\omega _I $
for different $q$ increase or decrease linearly with respect to
$v$. In general, as $q$ increases from 0 to 1, $\omega _R $
increases for different $v$. $\omega _I $ exhibits the behavior of
first decrease and latter increase with the increase of $q$ for
different $v$, which was also observed in the study of stationary
case [23].

\begin{figure}
\resizebox{0.5\linewidth}{!}{\includegraphics*{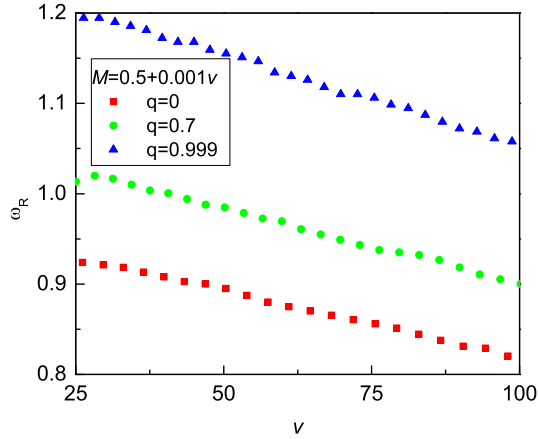}}
\caption{The frequency $\omega _R $ for $l = 2$, evaluated at $r =
5$ in the linearly absorbing black hole $M(v) = 0.5 + 0.001v$.}
\label{fig3}
\end{figure}

\begin{figure}
\resizebox{0.5\linewidth}{!}{\includegraphics*{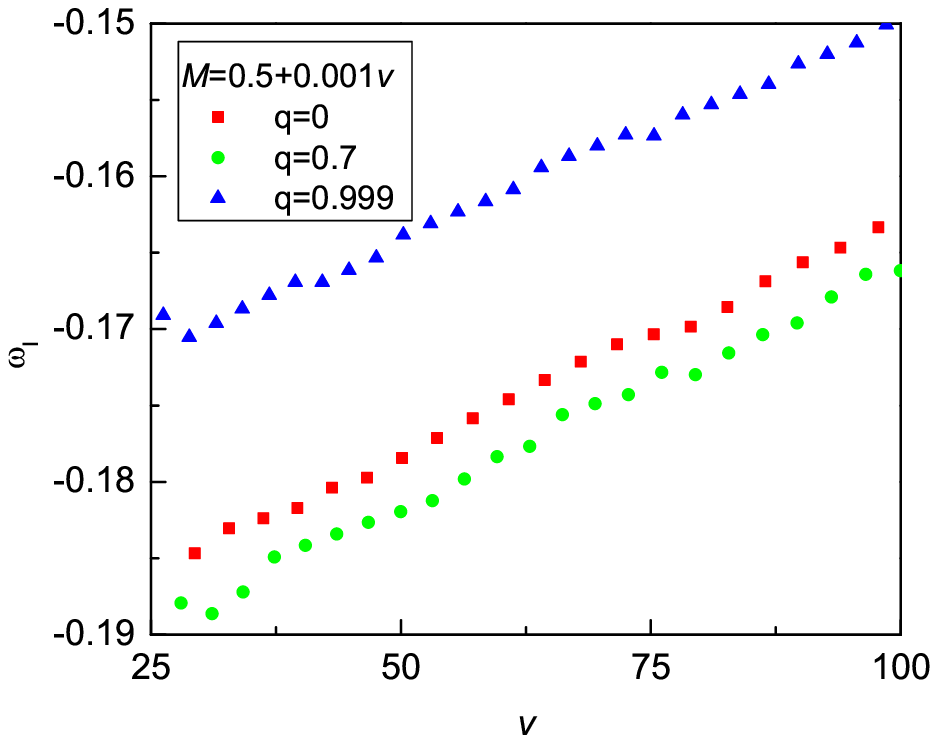}}
\caption{The frequency $\omega _I $ for $l = 2$, evaluated at $r =
5$ in the linearly absorbing black hole $M(v) = 0.5 + 0.001v$.}
\label{fig4}
\end{figure}

\begin{figure}
\resizebox{0.5\linewidth}{!}{\includegraphics*{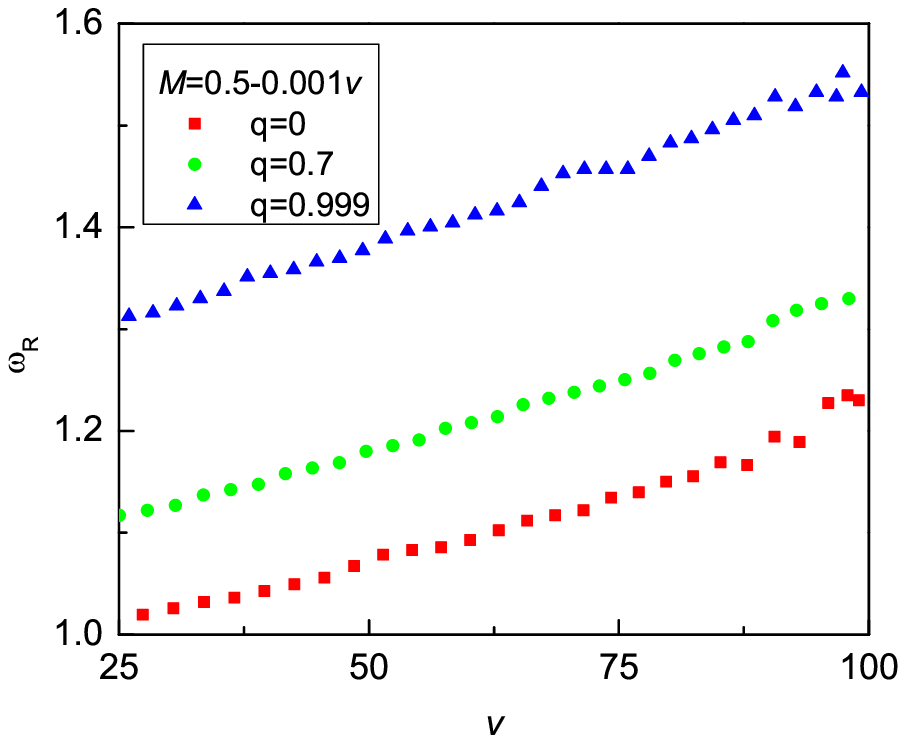}}
\caption{The frequency $\omega _R $ for $l = 2$, evaluated at $r =
5$ in the linearly evaporating black hole $M(v) = 0.5 - 0.001v$.}
\label{fig5}
\end{figure}

\begin{figure}
\resizebox{0.5\linewidth}{!}{\includegraphics*{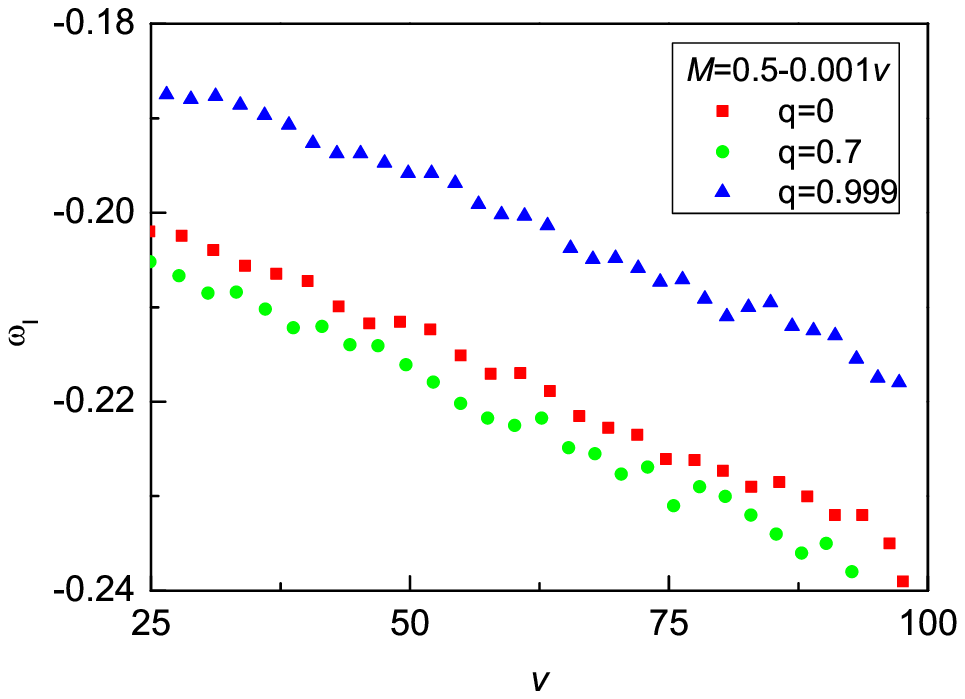}}
\caption{The frequency $\omega _I $ for $l = 2$, evaluated at $r =
5$ in the linearly evaporating black hole $M(v) = 0.5 - 0.001v$.}
\label{fig6}
\end{figure}

In order to compare with the QNMs in the stationary RN black hole,
we plot the curves $c_R (l,q) / \omega _R $ and $c_I (l,q) /
\omega _I $ with respect to time $v$ as shown in Fig.7-10. $c_R
(l,q)$ and $c_I (l,q)$ are defined in (\ref{eq11}). It is
interesting to note that all these curves for $M(v) = 0.5 +
0.001v$ are nearly equal for different $q$. Similar results are
obtained for the curves corresponding to $M(v) = 0.5 - 0.001v$.
The slope of the curve $c_R (l,q) / \omega _R $ (or $c_I (l,q) /
\omega _I )$ is equal to the $\dot {M}$, which suggests that the
formula (\ref{eq11}) is still a good approximation. In fact we
find
\begin{equation}
\label{eq29} M(v - v') \approx \frac{c_R (l,q)}{\omega _R (v-
v')},M(v - v') \approx \frac{c_I (l,q)}{\omega _I (v- v')} ,
\end{equation}
where the parameter $v'$ represents the retarded effect. As shown
in Fig.11, the frequencies $\omega _R $ for different position
radius $r$ with $l = 2$, $q = 0$ and $M(v) = 0.5 - 0.001v$ are
plotted. Clearly, the temporal evolution of the field $\Psi (r,v)$
at a radius $r = 20$ has a retard corresponding to the position $r
= 10$.

\begin{figure}
\resizebox{0.5\linewidth}{!}{\includegraphics*{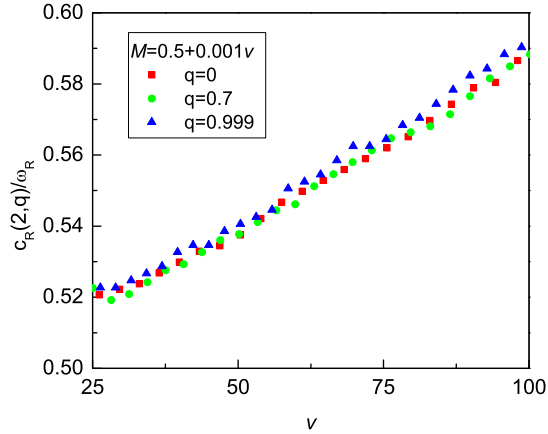}}
\caption{$c_R (l,q) / \omega _R $ as the function of time $v$ for
$l = 2$, $r = 5$ and $M(v) = 0.5 + 0.001v$.} \label{fig7}
\end{figure}

\begin{figure}
\resizebox{0.5\linewidth}{!}{\includegraphics*{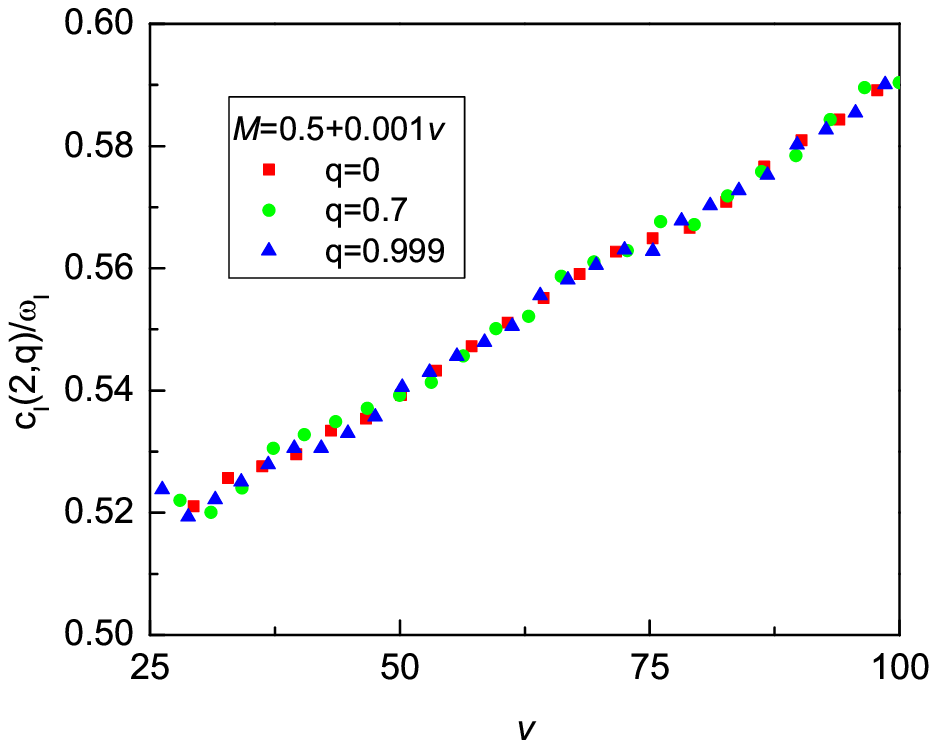}}
\caption{$c_I (l,q) / \omega _I $ as the function of time $v$ for
$l = 2$, $r = 5$ and $M(v) = 0.5 + 0.001v$.} \label{fig8}
\end{figure}

\begin{figure}
\resizebox{0.5\linewidth}{!}{\includegraphics*{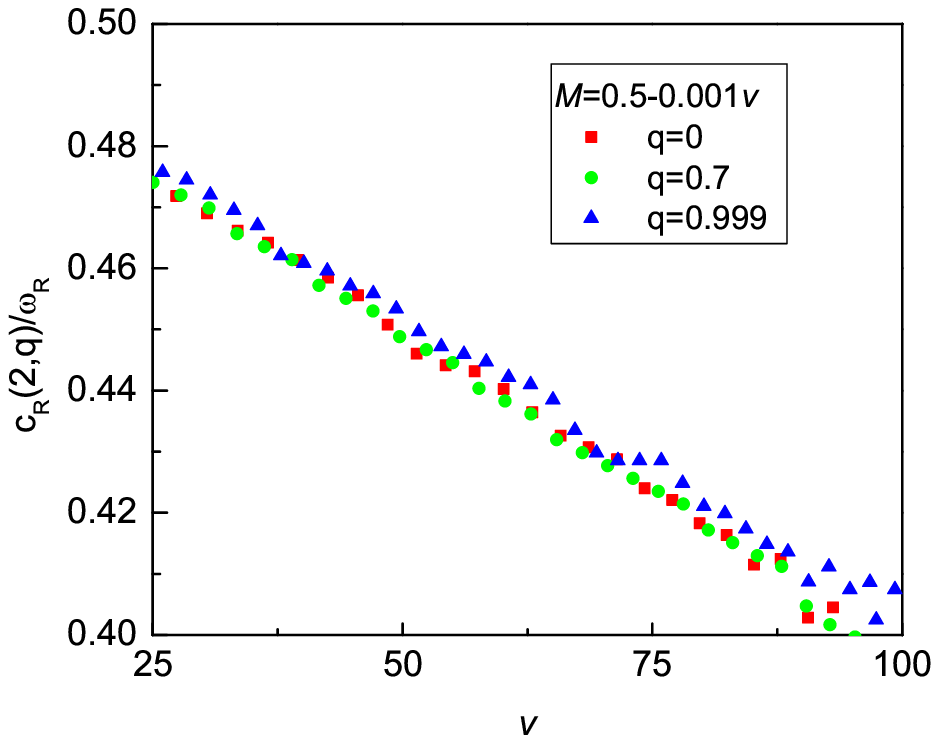}}
\caption{$c_R (l,q) / \omega _R $ as the function of time $v$ for
$l = 2$, $r = 5$ and $M(v) = 0.5 - 0.001v$. } \label{fig9}
\end{figure}

\begin{figure}
\resizebox{0.5\linewidth}{!}{\includegraphics*{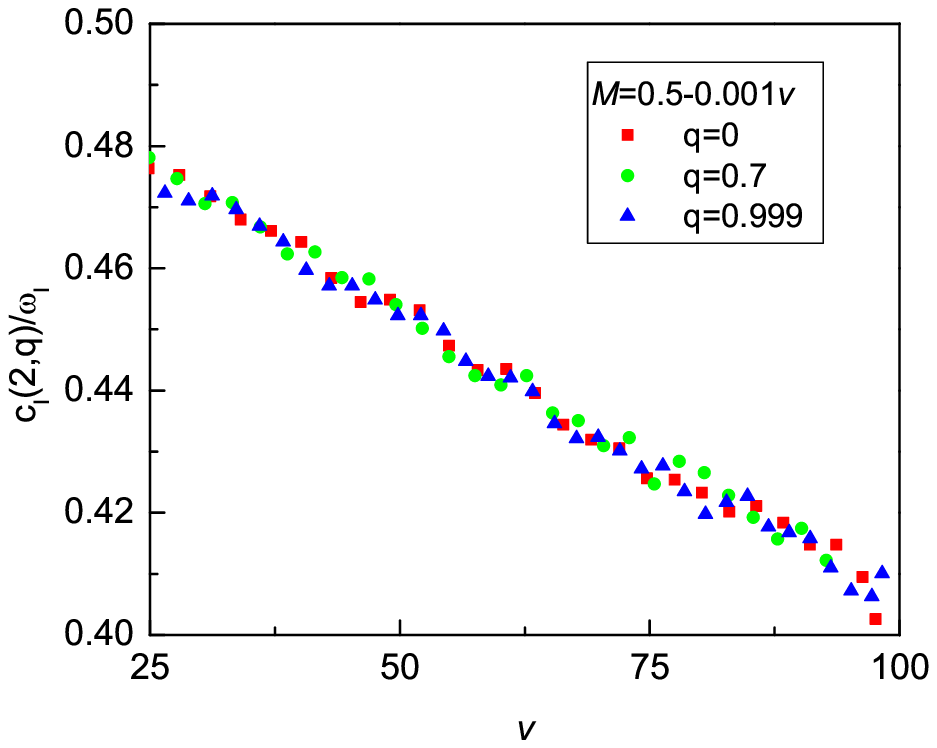}}
\caption{$c_I (l,q) / \omega _I $ as the function of time $v$ for
$l = 2$, $r = 5$ and $M(v) = 0.5 - 0.001v$.} \label{fig10}
\end{figure}

\begin{figure}
\resizebox{0.5\linewidth}{!}{\includegraphics*{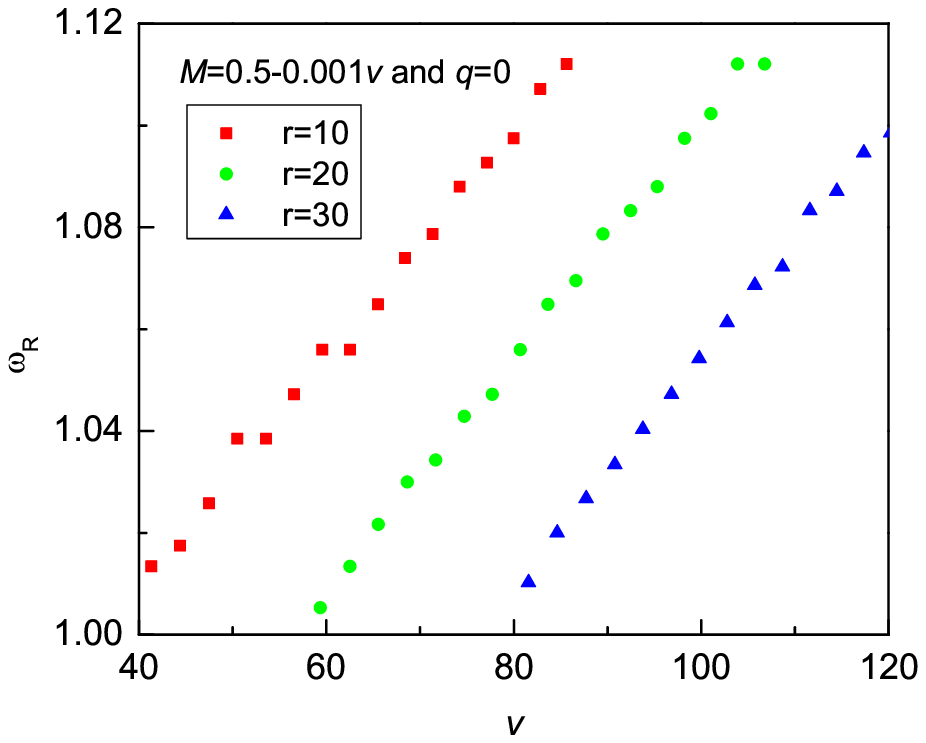}}
\caption{$\omega _R $ as the function of time $v$ for $l = 2$, $q
= 0$, $M(v) = 0.5 - 0.001v$, evaluated at different position $r =
10$, $r = 20$ and $r = 30$. The temporal evolution of the field
$\Psi (r,v)$ at radius $r = 20$ (or $r = 30)$ has a retard to the
position $r = 10$. } \label{fig11}
\end{figure}

Now we extend our discussion to the quick change of the black hole
mass. For the linear model we adopted is shown in Fig.12. With the
increase of the black hole mass, we find the same qualitative
behavior of the change of the QNM frequencies as that we observed
above.  The time scale of the variation of the black hole mass is
now shorter than that of the QNM frequencies. The observed
behavior of the quasinormal ringing is shown in Fig.13. $\omega _R
$ and $\omega _I $ are determined and exhibited in Fig.14, 15.  We
have shown that in the case when the variation timescale of black
hole mass is shorter than that the QNM frequency, the waveform and
the behavior of the QNM frequencies are the same as those we
discussed for the slowly changing of the black hole mass.

In the physical sense, the waveform and the frequencies of QNM's can be
understood in terms of wave scattering in a given spacetime [27] and
they carry significant information about the background. The amplitude of
the wave decreases with the number of scatterings. The repeated scatterings
will retard the influence on the temporal evolution of the wave by the
time-dependent potential due to the quick change of the black hole mass,
which leads the time scale for the change of the mass to be different
from the time scale for the change of the frequencies. In general case,
the repeated scatterings cause the time scale for the change of the QNM
frequencies to be longer than the time scale for the quick change of the
black hole mass.

\begin{figure}
\resizebox{0.5\linewidth}{!}{\includegraphics*{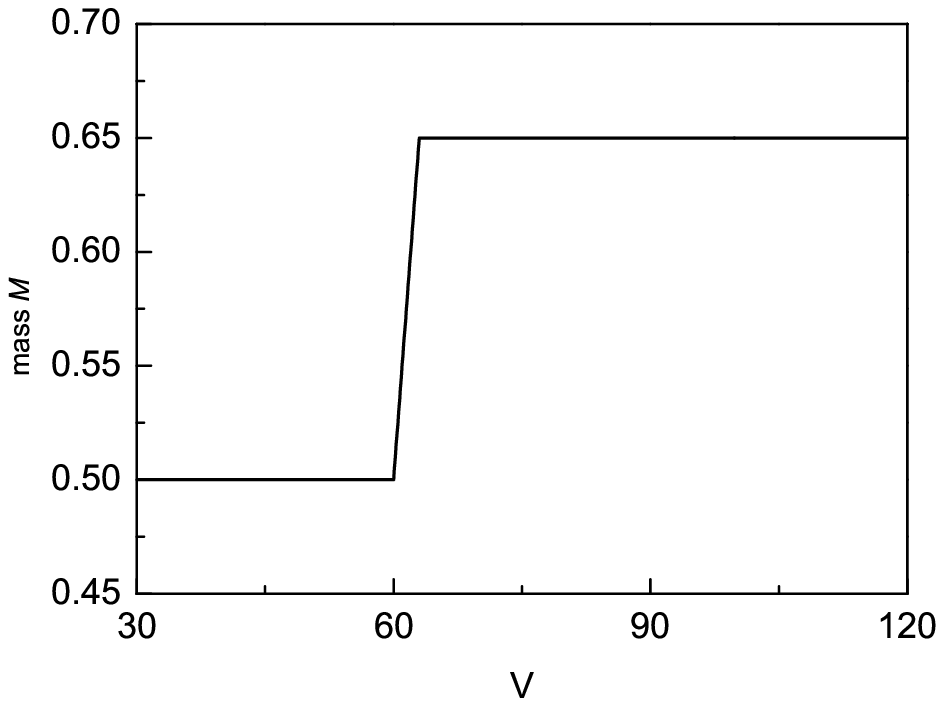}}
\caption{Considering the black hole mass:$q=0.999$, $M(v)=0.5$ for
$v<60$, $M(v)=0.5+0.05(v-60)$ for $60 \le v \le 63$ and
$M(v)=0.65$ for $v>60$. } \label{fig12}
\end{figure}

\begin{figure}
\resizebox{0.5\linewidth}{!}{\includegraphics*{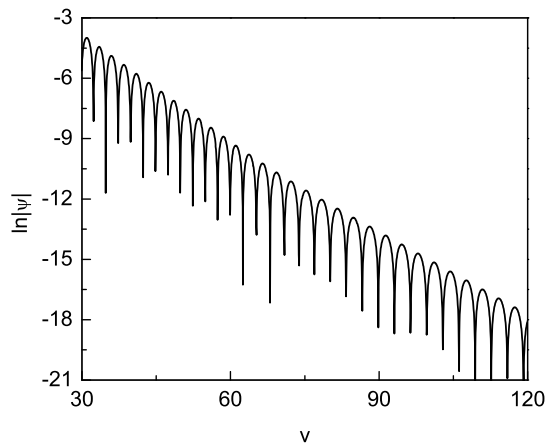}}
\caption{At $r=5$, the observed behavior of the quasinormal
ringing for $l=2$.} \label{fig13}
\end{figure}

\begin{figure}
\resizebox{0.5\linewidth}{!}{\includegraphics*{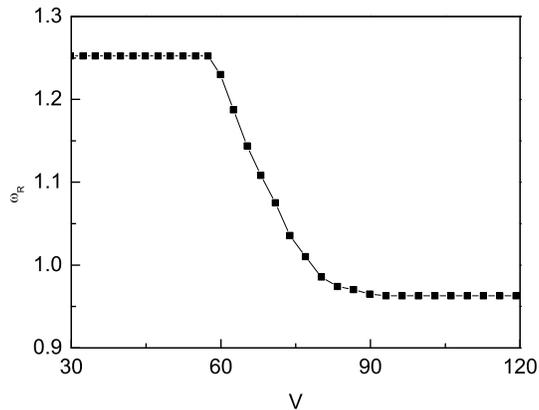}}
\caption{Change of the real part of the quasinormal frequency.
Black hole mass starts to increase when $v=60$, and the real part
of the QNM frequency starts to decrease at the same moment. It is
clear that time-scale of the change of the QNM frequency is much
longer than the change of the black hole mass($dv=3$).}
\label{fig14}
\end{figure}

\begin{figure}
\resizebox{0.5\linewidth}{!}{\includegraphics*{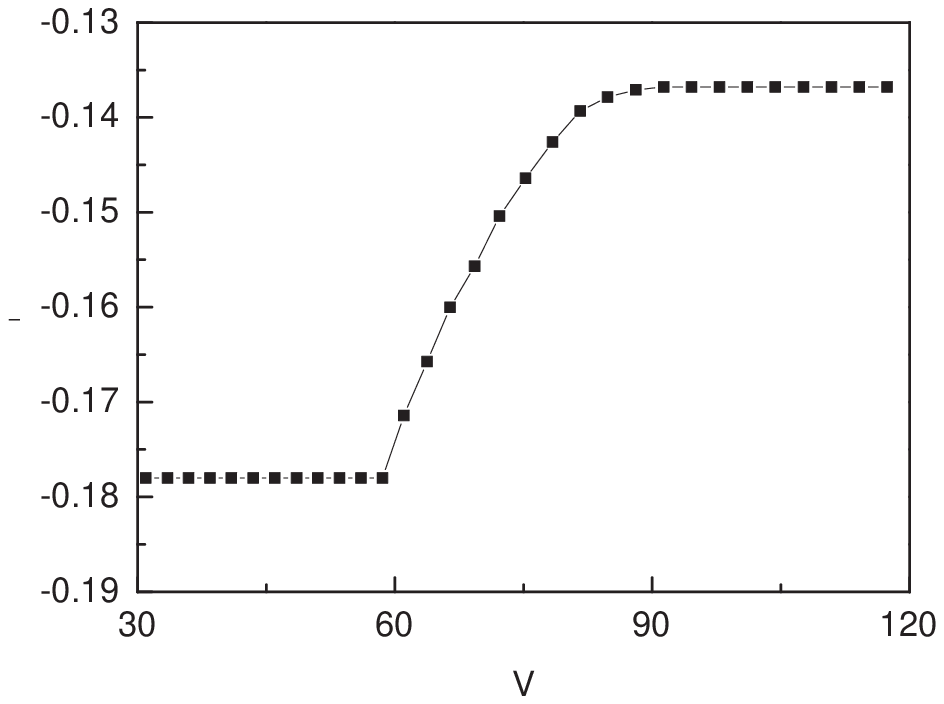}}
\caption{Change of the imaginary part of the QNM frequency. Black
hole mass starts to increase when $v=60$, and the imaginary part
of the QNM frequency starts to increase at the same moment. It is
clear that time-scale of the change of the QNM frequency is much
longer than the change of the black hole mass($dv=3$).}
\label{fig15}
\end{figure}

We have also extended our discussion to the model with
exponentially decreasing mass and charge [26],
\begin{equation}
\label{eq30}
\begin{array}{l}
 M(v) = \left\{ {{\begin{array}{l l}
 {m_0 } & {v \le v_0 } \\
 {m_0 \exp ( - \alpha v)} & {v_0 \le v \le v_1 } \\
 {m_1 } & {v \ge v_1 } \\
\end{array} }} \right. \\
 Q(v) = qM(v) ,\\
 \end{array}
\end{equation}
where $m_0 $, $m_1 $, $\alpha $ and $q$ are constant parameters.
Changing the sign before $\alpha$, (\ref{eq30}) can be used to
describe the absorption process. In order to compare the different
results between the exponential model and the linear model, we
assume that there are two absorbing black holes. The mass of one
black hole increases from $M(v_0 ) = 0.5$ to $M(v_1 ) = 1$
exponentially, while the other increases from $M(v_0 ) = 0.5$ to
$M(v_1 ) = 1$ linearly. Fig.16 shows the $c_R (l,q) / \omega _R $
and $c_I (l,q) / \omega _I $ as functions of time $v$ for the
linear and exponential model. The relative parameters in both
models are selected as $l = 2$, $r = 5$, $q = 0.999$, $v_0 = 0$
and $v_1 = 150$. The two dot lines are the mass functions $M(v -
{v}')$ for models (\ref{eq30}) and (\ref{eq26}) with ${v}' = 8$.
In Fig.17 we plotted $c_R (l,q) / M\omega _R $ and $c_I (l,q) /
M\omega _I $ as a means to check the validity of (\ref{eq29}). The
behaviors we have observed in Figs. 16, 17 also hold for other
values of $q$ (see Fig. 18 and Fig. 19 for $q$=0 as an example).
In the linear model both $1 / \omega _R $ and $1 / \omega _I $
depend linearly on $v$, while in the exponential model both $1 /
\omega _R $ and $1 / \omega _I $ depend exponentially on $v$. The
consistency of $M(v - v')$ and $c_R (l,q) / \omega _R $ (or $c_I
(l,q) / \omega _I )$ implies the validity of the equations
(\ref{eq29}). We have also checked the validity of the equations
(\ref{eq29}) for different $l$, such as $l = 3,4,5$. The result
keeps the same when the variation time scale of the black hole
mass is shorter than that of the QNM frequency.

\begin{figure}
\resizebox{0.5\linewidth}{!}{\includegraphics*{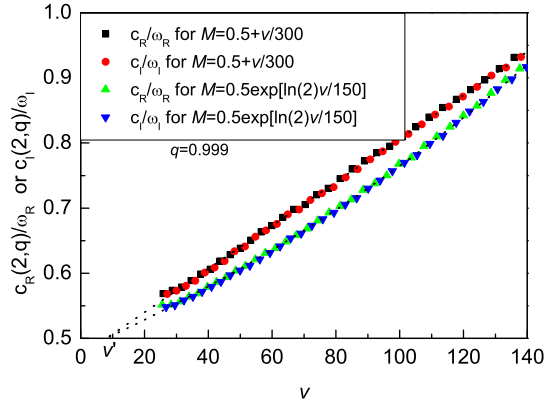}}
\caption{$c_R (l,q) / \omega _R $ and $c_I (l,q) / \omega _I $ as
the functions of time $v$ for the linear and exponential model. In
the linear model, $\lambda = - 1 / 150$, and in the exponential
model $\alpha = - (\ln 2) / 150$. Both models describe the
absorption process of the black hole, where the mass increases
from $M(v_0 ) = 0.5$ to $M(v_1 ) = 1$. The relative parameters in
both models are selected as $l = 2$, $r = 5$, $q = 0.999$, $v_0 =
0$ and $v_1 = 150$. The two dot lines are the mass functions $M(v
- v')$ for models (\ref{eq30}) and (\ref{eq26}) with ${v}' = 8$.}
\label{fig16}
\end{figure}

\begin{figure}
\resizebox{0.5\linewidth}{!}{\includegraphics*{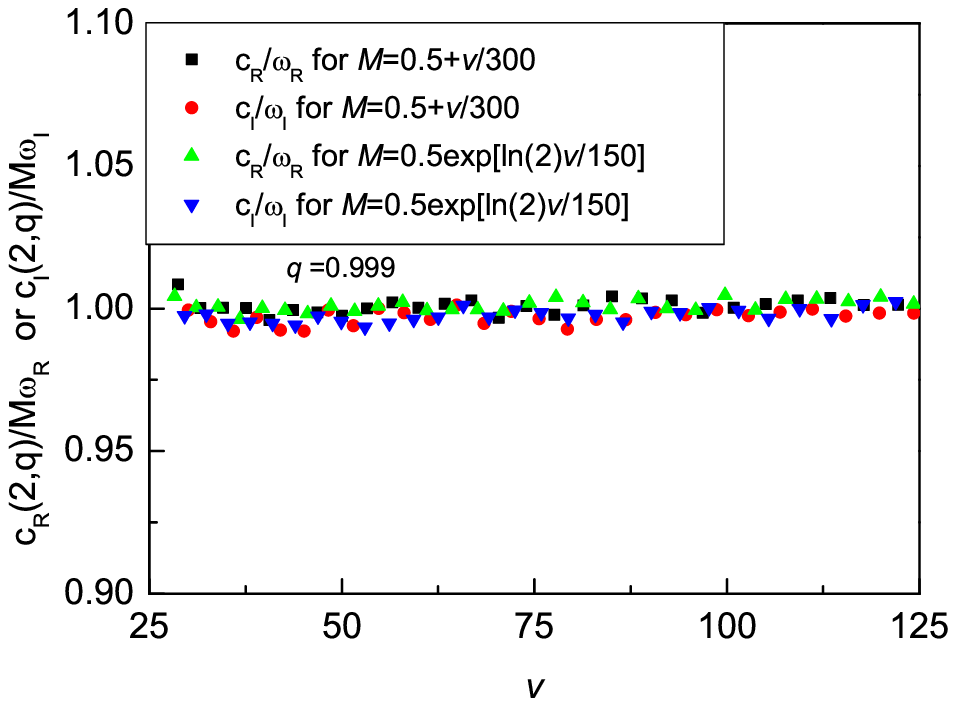}}
\caption{$c_R (l,q) / M\omega _R $ and $c_I (l,q) / M\omega _I $
as the functions of time $v$. The corresponding parameters are the
same as chosen in Fig. 16.} \label{fig17}
\end{figure}

\begin{figure}
\resizebox{0.5\linewidth}{!}{\includegraphics*{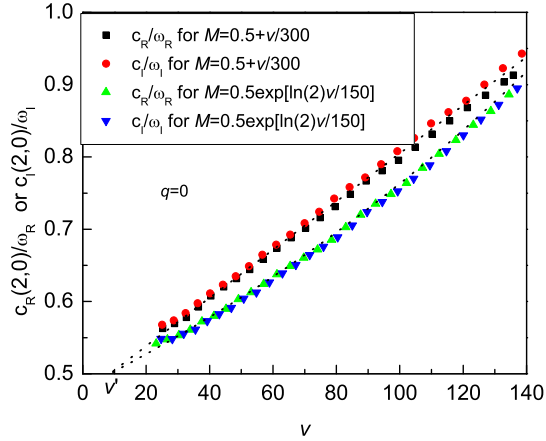}}
\caption{$c_R (l,q) / \omega _R $ and $c_I (l,q) / \omega _I $ as
the functions of time $v$ in the case of $q$=0. The corresponding
parameters are the same as chosen in Fig. 16.} \label{fig18}
\end{figure}

\begin{figure}
\resizebox{0.5\linewidth}{!}{\includegraphics*{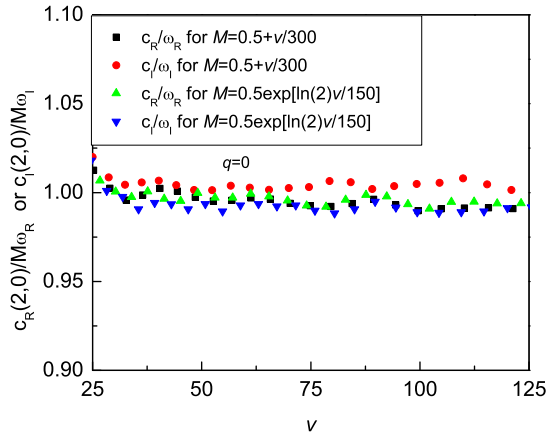}}
\caption{$c_R (l,q) / M\omega _R $ and $c_I (l,q) / M\omega _I $
as the functions of time $v$. The corresponding parameters are the
same as chosen in Fig. 16.} \label{fig19}
\end{figure}

\section{Summary and discussion}

We have studied the evolution of the massless scalar field
propagating in a time-dependent charged Vaidya black hole
background. The two generalized tortoise coordinate
transformations were used to study the evolution of the massless
scalar field. In the first case we used both the $r_ + $ and  $r_
- $, which is appropriated for the case of large charge in
numerical calculation. In the other case we only used the $r_ + $,
which is appropriated for the case of small charge. The wave
equation with a time-dependent scattering potential is derived. In
our numerical study, we have found the modification of the QNMs
due to the temporal dependence of the black hole spacetimes. In
the absorption process, when the black hole becomes bigger, both
the imaginary frequency $\left| {\omega _I } \right|$ and the real
frequency $\omega _R $ decrease with time. However, in the
evaporating process, when the black hole loses mass, both $\left|
{\omega _I } \right|$ and $\omega _R $ increase with the increase
of time. The study shows that, for the slowest damped QNMs, the
approximate formulas in stationary RN black hole is a good
description for the time-dependent charged Vaidya black hole,
which can be expressed as (\ref{eq29}).

In general, we conclude that the time dependent processes are well
described by an adiabatic approximation with a delay in the
definition of the mass function, borrowing the results at each
moment from the problem of a black hole of a time-independent
masses, instantly equal to the map of the equivalent Vaidya black
hole at a retarded time. This implies a huge simplification for
the problem of determination of signals from time-dependent
processes near black hole. Finally we notice that the period
corresponding to QNM is much smaller than the time of flight of a
relativistic object in the black hole, actually by several orders
of magnitude. We thus conclude that the value we used for the
$\lambda $ parameter is large enough to include realistic cases.

\begin{acknowledgments}
This work was partially supported by  NNSF of China, Ministry of
Education of China and Shanghai Science and Technology Commission.
E. Abdalla's work was partially supported by FAPESP and CNPQ,
Brazil. R.K. Su's work was partially supported by the National
Basic Research Project of China.
\end{acknowledgments}


\end{document}